\documentclass[12pt,twoside,letterpaper,fleqn]{article}
\ifx\pdfoutput\undefined
\usepackage[pdftex,dvips,bookmarks,breaklinks]{hyperref}
\else
\usepackage{hyperref}
\fi
\hypersetup{colorlinks=false,bookmarksopen,bookmarksnumbered,citecolor=blue}

\usepackage{latexsym}

\usepackage{authblk}

\usepackage{amssymb,amsfonts,amsmath}
\usepackage{graphicx}
\usepackage{indentfirst}
\usepackage{bbm}
\usepackage{color}

\parskip=\medskipamount

\def\Real{{\mathbb R}}

\def\1{1\hspace{-4pt}1}
\def\j1{\widetilde{1\hspace{-4pt}1}}

\def\bec{\begin{center}}
\def\ec{\end{center}}
\def\a{\alpha}

\def\b{\beta}

\def\c{\gamma} 
\def\C{\Gamma}
\def\d{\delta} 

\def\e{\epsilon}

\def\r{\rho}

\def\th{\theta}

\def\o{\omega}

\def\nn{\nonumber}
\newcommand{\eq}[1]{({#1})}

\def\ed{\end{document}}

\def\be{\begin{equation}}
\def\ee{\end{equation}}
\def\bea{\begin{eqnarray}}
\def\eea{\end{eqnarray}}
\def\ba{\begin{array}}
\def\ea{\end{array}}

\begin{document}
\title{2D Poisson Sigma Models with \\[4pt]Gauged Vectorial Supersymmetry}

\author[a,b]{Roberto Bonezzi \thanks{roberto.bonezzi@bo.infn.it}}
\author[b]{Per Sundell\thanks{per.anders.sundell@gmail.com}}
\author[b,c]{Alexander Torres-Gomez\thanks{alexander.torres.gomez@gmail.com}}
\affil[a]{\textit{\small Dipartimento di Fisica ed Astronomia, Universit\`a di Bologna and
INFN, Sezione di Bologna,\qquad via Irnerio 46, I-40126 Bologna, Italy}}
\affil[b]{\textit{\small Departamento de Ciencias F\'isicas, Universidad Andres Bello, Republica 220, Santiago, Chile}}
\affil[c]{\textit{\small Instituto de Ciencias F\'isicas y Matem\'aticas, Universidad Austral de Chile-UACh, Valdivia, Chile}}

\renewcommand\Authands{  and }

\date{}
\maketitle
\thispagestyle{empty}

\abstract{\noindent In this note, we gauge the rigid vectorial supersymmetry of the two-dimensional Poisson sigma model presented in arXiv:1503.05625.
We show that the consistency of the construction does not impose any further constraints on the differential Poisson algebra geometry than those required for the ungauged model.
We conclude by proposing that the gauged model provides a first-quantized framework for higher spin gravity.}

\newpage
\setcounter{page}{1}
\tableofcontents
\section{Introduction}

\noindent The original two-dimensional Poisson sigma model \cite{Ikeda,Schaller}
provide a quantum field theoretic implementation \cite{CaFe} of Kontsevich's formality theorem \cite{Kontsevich},
which states that, modulo the issue of ordering prescriptions, there is one-to-one correspondence between algebras of functions equipped with Poisson brackets and nonlocal products that are noncommutative and associative, known as star products.
More precisely, the Feynman diagrammatic expansion of the Poisson sigma model \cite{CaFe} reproduces Kontsevich's star product in the case of Poisson manifolds of $\Real^n$ topology.

It is natural to ask whether there exists an extension of the star product to differential forms that is manifestly covariant and compatible with a nilpotent operator that deforms the de Rham differential, referred to as differential star product algebras.
Indeed, at the semi-classical level, the Poisson algebras of functions can be extended to differential forms that are compatible with the de Rham differential \cite{Chu,Beggs}, also known as differential Poisson algebras.
Following the algebraic approach, these algebras have been shown \cite{Tagliaferro,Zumino,Finnish} to admit deformations that yield differential star product algebras up to second order in the deformation parameter, except in the torsion free symplectic case where the star product is given by a covariantized Moyal formula to all orders.

In \cite{us}, we have constructed a natural extension of the original Poisson sigma model by including vectorial fermions, and proposed that its perturbative quantization yields the full differential star product algebra on a manifestly covariant format.
A key feature of this model is that, besides its gauge symmetries, it exhibits a rigid nilpotent supersymmetry, \emph{i.e.} a symmetry generated by a holonomic vector field, that is the vectorial superspace counter part of the de Rham differential on
the original bosonic Poisson manifold.

In this technical note, we show that it is possible to gauge this holonomic symmetry,
following the standard Noether approach, without the need to impose any further constraints on the differential Poisson algebra geometry.
This construction opens up a framework for constructing topological open string fields theories within the framework of Cartan integrable systems.
In particular, based on \cite{GR_as_CS,FCSDD,FCS4D}, we propose that models
of these type may provide first-quantized formulations of higher spin gravities,
as we shall comment on further in the Conclusions.

The paper is organized as follows: Section \ref{sec:2} describes the basic features of the ungauged model and its proposed usage in deformation quantization.
The gauged model is formulated and analyzed in Section \ref{sec:3}, with particular attention paid to its universal Cartan integrability.
We conclude with a proposal for higher spin gravity in Section \ref{sec:4}.
Appendix \ref{app:a} contains some useful identities for tensor calculus
on Poisson manifolds.

\section{The ungauged model}\label{sec:2}

\subsection{General framework}

\noindent While we are here going to focus on a particular class of
two-dimensional models, related to associative free differential
algebras, these models are part of a larger framework for
constructing homotopy associative quantum algebras with free
differentials based on topological open p-branes
\cite{Park,Ikeda2001,Hofman1, Hofman2}.
The basic idea is to use perturbative quantization on the
worldvolumes to deform the semi-classical algebras of differential
forms $\Lambda(N)$ on graded target space manifolds $N$.
As proposed in \cite{us}, in order to include elements in $\Lambda(N)$
of arbitrary degrees, the manifold $N$ should be extended
to a vectorial superspace $T_{\rm f}N$, in the sense that
for each coordinate of $N$, $X^\a$ say, of degree $p_\a$,
one introduces a fermionic partner, $\Theta^\a$, of the same
degree.
Thus, the bundle $T_{\rm f}N$ has local vectorial
supercoordinates $(X^\a,\Theta^\a)$ with bi-grading
\be {\rm deg}(X^\a,\Theta^\a)=(p_\a,p_\a)\ ,\qquad
\e_{\rm f}(X^\a,\Theta^\a)=(0,1)\ .\ee
As for the rest of the construction, following the usual
procedure, additional momentum variables, say $(P_\a,\lambda_\a)$,
are introduced, with
\be {\rm deg}(P_\a,\lambda_\a)=(p-p_\a,p-p_\a)\ ,\qquad
\e_{\rm f}(P_\a,\lambda_\a)=(0,1)\ .\ee
playing the role of Lagrange multipliers on the worldvolume.
Thus, the sigma model map $\varphi: {\cal M}_{p+1}\rightarrow {\cal N}$
sends the worldvolume to a target bundle ${\cal N}$
over $T_{\rm f}N$ that is $(\mathbb N,\mathbb Z_2)$-graded,
in the sense that all tensorial objects ${\cal O}$ over ${\cal N}$ have
a degree ${\rm deg}({\cal O})$ and an additional
Grassmann parity $\e_{\rm f}({\cal O})$.
Finally, in order to have well-defined semi-classical
algebra the Koszul sign convention is taken to be
\be {\cal O}{\cal O}'=(-1)^{{\rm deg}({\cal O}){\rm deg}({\cal O}')+
\e_{\rm f}({\cal O})\e_{\rm f}({\cal O}')} {\cal O}'{\cal O}\ .\ee
The classical action of the extended model with fermions
takes the same form as that of the original bosonic model,
\emph{viz.}
\be S_{\rm cl}=\int_{{\cal M}_{p+1}}\varphi^\ast(\vartheta-{\cal H})\ ,\ee
where $\vartheta$ is the presymplectic form on ${\cal N}$
and ${\cal H}$ is a generalized Hamiltonian obeying $\{{\cal H},{\cal H}\}=0$,
using the bracket defined by the symplectic form $\Omega=d\vartheta$.
By means of the AKSZ procedure \cite{AKSZ}, the classical action
can then be extended further to a Batalin--Vilkovisky \cite{BV1,BV2,BV3}
(quantum) master action with AKSZ superfields with distinct
total degree $|\cdot|$, equal to the sum of ghost number and
(classical form) degree, and Grassmann parity; the Koszul sign
convention now takes the form
\be {\cal O}{\cal O}'=(-1)^{|{\cal O}| |{\cal O}'|+
\e_{\rm f}({\cal O})\e_{\rm f}({\cal O}')} {\cal O}'{\cal O}\ .\ee
In particular, working on ${\cal N}$, the differential $d$ on $N$,
which has
\be {\rm deg}(d)=1\ ,\qquad \e_{\rm f}(d)=0\ ,\ee
can be realized as a holonomic vector field $\delta_{\rm f}$ on $T_{\rm f}N$,
which has
\be {\rm deg}(\delta_{\rm f})=0\ ,\qquad \e_{\rm f}(\delta_{\rm f})=1\ ,\ee
by mapping $p$-forms $\omega$ on $N$ to zero-forms $f_\omega$ on $T_{\rm f}N$
that are $p$-linear functions of the fermionic fiber coordinates, \emph{viz.}
\be \omega=\tfrac1{p!} dX^{\a_1}\wedge\cdots\wedge dX^{\a_p} \omega_{\a_1\cdots\a_p}\ \mapsto\ f_\omega=\tfrac1{p!} \Theta^{\a_1}\cdots \Theta^{\a_p}\; \omega_{\a_1\cdots\a_p}\ ,\label{mapsto}\ee
and taking
\be \delta_{\rm f} X^\a=\Theta^\a\ ,\qquad \delta_{\rm f}\Theta^\a=0\ .
\label{deltafX}\ee

\subsection{Differential Poisson algebras}

\noindent As for the underlying target space geometry, upon going to
canonical coordinates, the expansion of the
Hamiltonian in momenta yields a set of mutually
compatible poly-vector fields and related geometric
structures on $N$ (including pre-connections), that constitute
graded (anti-)symmetric brackets obeying natural graded
generalizations of Leibniz' rule and the Jacobi identity,
defining a generalized differential Poisson algebra on $N$.
The deformation quantization of such brackets can be achieved
following a purely algebraic approach.
In particular, the algebra of zero-forms on a manifold $N$
of degree $0$ equipped with an ordinary bi-vector field $\Pi=\Pi^{\a\b}\partial_\a\wedge \partial_\beta$ obeying the Poisson condition $\{\Pi,\Pi\}_{\rm Schouten}=0$, such that the Poisson bracket $\{f,g\}=\Pi(df,dg)$ between zero-forms obeys the Jacobi identity,
can be deformed to an associative algebra equipped with Kontsevich's star product.

The semi-classical Poisson bracket algebra can be extended to a differential Poisson algebra with a Poisson bracket between general differential forms that is compatible with the de Rham differential.
This extension can be achieved by introducing a connection one-form $\Gamma^\a{}_\b=d\phi^\c\Gamma^\a_{\c\b}$ and an additional
tensorial one-form $S=d\phi^\a S^{\b\c}_\a \partial_\b\odot \partial_\c$,
such that the bracket between a function $f$ and a one-form $\omega$ reads
\be \{f,\omega\}=\Pi^{\a\b}\partial_\a f \nabla_\beta \omega+ (\widetilde \nabla \Pi^{\a\b}+S^{\a\b})\partial_\a f \imath_\b \omega\ ,\label{S}\ee
where $\widetilde\nabla$ is defined using the connection one-form
$\widetilde\Gamma^\a{}_\b=d\phi^\c\widetilde\Gamma^\a_{\c\b}$ where
$\widetilde\Gamma^\a_{\c\b}=\Gamma^\a_{\c\b}+T^\a_{\b\c}$, \emph{i.e.}
$\widetilde\nabla\omega=\nabla\omega+2\imath_T \omega$ where $T$ is treated
as a vector-valued differential two-form.
It is convenient to work with a connection that
annihilates the Poisson structure, that is, to make
the following compatibility assumption:
\be \widetilde \nabla_\alpha \Pi^{\beta \gamma}=0\ .\label{comp1}\ee
Its integrability requires that the curvature two-form
$\widetilde R^\a{}_\b=\tfrac12 d\phi^\gamma\wedge
d\phi^{\d} \widetilde R_{\gamma\d}{}^{\a}{}_\b$ obeys
\be
\widetilde R^{\a\b}:= \Pi^{\beta \gamma} \widetilde R^\a{}_\gamma = \widetilde R^{\b\a}\ ,
\label{comp2}\ee
and, as shown in the Appendix, that
\begin{equation}\label{Pi(R-R-NT)}
\Pi^{\lambda [\beta}\left( R_{\lambda \epsilon}{}^{\gamma]}{}_\alpha+ R_{\alpha \lambda}{}^{\gamma]}{}_\epsilon- \nabla_\lambda T^{\gamma]}_{\epsilon \alpha} \right) =0 \ .
\end{equation}
Assuming in addition that $S=0$, the resulting Poisson bracket between two differential forms $\omega$ and $\eta$ can be written as
\be \{\omega,\eta\}=\Pi^{\a\b} \nabla_\a \omega\wedge \nabla_\b \eta+
(-1)^{{\rm deg}(\omega)} \widetilde R^{\a\b} \imath_\a \omega\wedge
\imath_\b \eta\ ,\label{PB}\ee
where the last term thus ensures that $d\{\omega,\eta\}=\{d\omega,\eta\}+(-1)^{{\rm deg}(\omega)} \{\omega,d\eta\}$.
As has been shown in \cite{Chu, Beggs, Tagliaferro, Zumino, Finnish}, the graded Jacobi identities hold provided that the torsion and the curvature obey integrability conditions, listed below in Eqs. \eqref{J000}--\eqref{J111}.

We would like to remark that whereas there is no obstruction to deforming a Poisson bracket into a star product in a space of zero-forms, which is one of the main results of Kontsevich's formality theorem, there is to our best understanding no classification of the obstructions, if any, to deforming a differential Poisson algebra into a differential star product algebra.

\subsection{2D Poisson sigma model with holonomic symmetry}

Returning to the field theoretic point-of-view, the natural approach towards the deformation quantization of the aforementioned bracket structures thus consists of coupling the underlying integrable geometric structures to open topological p-branes, as outlined above.
One can then treat the classical differential forms as symbols for boundary vertex operators, with the aim of obtaining the differential star product algebra  perturbatively using the AKSZ formalism, which we leave for future work.

As far as the deformation of standard Poisson brackets between zero-forms is concerned, the perturbative quantization of the two-dimensional Ikeda--Schaller--Strobl Poisson sigma model \cite{Ikeda,Schaller} was shown by Cattaneo and Felder \cite{CaFe} to re-produce Konsevich's explicit star product formula on $N=\Real^n$ \cite{Kontsevich}, indeed without encountering any obstructions albeit on a non-manifestly covariant form.

More recently, in \cite{us}, we have proposed that the manifestly covariant deformation quantization of the differential Poisson algebra, including differential forms in higher degrees, can be achieved starting from the aforementioned natural extension of the Ikeda--Schaller--Strobl model, which thus facilitates the mapping \eqref{mapsto} of  line-elements $d\phi^\a$ on $N$, with degree one and fermion number zero, to the fiber coordinates $\theta^\alpha$ on $T_{\rm f}N$, with degree zero and fermion number one.

To write the model down, one thus introduces momenta $\eta_\a$ and $\chi_\a$ for $\phi^\a$ and $\theta^\a$, which thus have degrees one and fermion numbers
zero and one, respectively.
The classical action is now given by \cite{us}
\begin{eqnarray}
S_0&=&\int_{M_2} \left(
\eta_\a \wedge d\phi^\a+\tfrac12 \Pi^{\a\b} \eta_\a \wedge \eta_\b
+\chi_\a \wedge \nabla \theta^\a+ \tfrac14 \widetilde R_{\gamma\d}{}^{\a\b} \chi_\a \wedge \chi_\b \theta^\gamma\theta^\d \right)\ ,
\end{eqnarray}
where the field variables $(\phi^\a, \eta_\a; \theta^\a,\chi_\a)$ are assigned form degrees ${\rm deg}_2(\cdot)$ on $M_2$ and an additional Grassmann parity $\e_{\rm f}(\cdot)$ as follows:
\begin{center}
\begin{tabular}{| l |c | c | c | c | } \hline
 &  $\phi^\alpha$ & $\eta_ \alpha$ & $\theta^\alpha$ & $\chi^\alpha$ \\ \hline
 $\text{deg}_2$ &  0 & 1 & 0 & 1  \\ \hline
 $\epsilon_{\text f}$ & 0 & 0 & 1 & 1\\ \hline
\end{tabular}
\end{center}
In the target space, the fields $\phi^\a$ are local coordinates of the base manifold $N$ of a bundle $T^\ast[1,0]N \oplus T^\ast[1,1]N\oplus T[0,1]N$ with fiber coordinates $(\eta_\a,\chi_\a,\theta^\a)$, where $T^\ast[m,\e]N$ is obtained from $T^\ast N\equiv
T^\ast[0,0]N$ by replacing the fiber coordinates by new fiber coordinates with degree $m$ and Grassmann parity $\e$, \emph{idem} $T[m,\e]N$.
The sigma model map $\varphi:M_2\rightarrow N$ induces a bundle over $M_2$ as follows:
\bea &&\varphi^\ast\Big(T^\ast[1,0]N \oplus T^\ast[1,1]N\oplus T[0,1]N\Big)\nn\\[5pt]
&=&\Big(\varphi^\ast(T^\ast[0,0]N)\otimes T^\ast M_2\Big) \oplus
\Big(\varphi^\ast(T^\ast[0,1]N)\otimes T^\ast M_2 \Big)\oplus
\varphi^\ast(T[0,1]N)\ ,\eea
whose sections we denote by $(\eta_\a,\chi_\a,\theta^\a)$ as well, for
simplicity of notation.
The covariant derivatives on $M_2$ of these sections are given by
\begin{equation}
\begin{split}
\nabla \theta^\alpha&=d\theta^\alpha+ d\phi^\beta \Gamma^\alpha_{\beta \gamma} \theta^\gamma \ , \\
\nabla \eta_\alpha&=d\eta_\alpha - d\phi^\gamma \Gamma^\beta_{\gamma \alpha} \eta_\beta \ , \\
\end{split}
\end{equation}
\emph{idem} $\chi_\alpha$. As shown in \cite{us}, the equations of motion,
which read
\begin{eqnarray}
{\cal R}^{\phi^\a}_0 &:=& d\phi^\a+ \Pi^{\a\b}\eta_\b=0\ ,\label{Rphi0}\\[5pt]
{\cal R}^{\theta^\a}_0 &:=& \nabla \theta^\a+ \tfrac12 \widetilde R_{\gamma\d}{}^{\a\b}\chi_\b\theta^\gamma \theta^\d =0\ ,\label{Rtheta0}\\[5pt]
{\cal R}^{\eta_\a}_0 &:=& \nabla \eta_\a +\tfrac 1 4  \nabla_\a \widetilde R_{\d\e}{}^{\b\gamma}\chi_\b \wedge \chi_\gamma  \theta^\d\theta^\e+ R_{\a \gamma}{}^{\b}{}_{\d}\chi_\b \wedge d\phi^\gamma \theta^\d = 0  \label{Reta0}\\[5pt]
{\cal R}^{\chi_\a}_0 &:=& \nabla \chi_\a - \tfrac12  \widetilde R_{\a\d}{}^{\b\gamma}\chi_\b  \wedge \chi_\gamma \theta^\d = 0\ ,\label{Rchi0}
\end{eqnarray}
form a universally Cartan integrable system provided that the background fields obey the conditions
\bea
&&\Pi^{\d [\a}T^\b_{\d\e} \Pi^{\gamma]\e} =  0   \ ,\label{J000}\\[5pt]
&& \Pi^{\alpha \rho} \Pi^{\sigma \beta} R_{\rho \sigma}{}^{\gamma}{}_\delta = 0 \ ,
\label{J001}\\[5pt]
&& \Pi^{\alpha \lambda} \nabla_\lambda \widetilde R_{\beta \gamma}{}^{\rho \sigma} = 0 \ ,\label{J011}\\[5pt]
&&\widetilde R_{\epsilon [ \rho}{}^{(\alpha \beta} \widetilde R_{\sigma \lambda]}{}^{\gamma) \epsilon} = 0\ ,\label{J111}
\eea
which in their turn are equivalent to that the Poisson bracket \eqref{PB} obeys the graded Jacobi identities.
More precisely, the integrability of the constraint on ${\cal R}^{\phi^\a}$ requires Eqs. \eqref{J000}--\eqref{J011}, while the integrability of the remaining constraints requires  Eqs. \eqref{J001}--\eqref{J111}.
In fact, decomposing the constraints into Young tables (not subtracting any traces), one finds that the covariant derivatives of \eqref{J000} and \eqref{J001} contain parts of \eqref{J001} and \eqref{J011}, respectively, while all of \eqref{J111} is contained in the covariant derivative of \eqref{J011}.

The compatibility of the bracket \eq{PB} with the exterior derivative
translates into the fact that the action has a rigid holonomic symmetry,
denoted by $\delta_\text f$, which acts on the fields as follows:
\bea\label{susy}
\delta_{\text f} \phi^\alpha&=&\theta^\alpha \ , \notag \\
\delta_{\text f} \theta^\alpha&=&0 \ , \notag \\
\delta_{\text f} \eta_\alpha&=&\Gamma^\beta_{\alpha\gamma} \eta_\beta \theta^\gamma+\tfrac1 2 \widetilde R_{\beta\gamma}{}^\delta{}_\alpha\, \chi_\delta \, \theta^\beta \theta^\gamma \ , \notag \\
\delta_{\text f} \chi_\alpha&=&-\eta_\alpha-\Gamma^\beta_{\alpha\gamma} \chi_\beta  \theta^\gamma \ .
\eea
The supersymmetry invariance of the action can be made manifest
by re-writing its Lagrangian on the following $\delta_\text f$-exact form:
\begin{equation}
S_0= \int_{M_2} \delta_{\text f} V\ ,\label{V}
\end{equation}
where the functional $V$, which is a fermionic two-form, is given by
\begin{equation}
V=\ -\chi_\alpha \wedge \left( d\phi^\alpha+\tfrac 1 2 \Pi^{\alpha \beta} \eta_\beta  \right)  \ ,
\end{equation}
where we have used \eqref{comp1}.
The corresponding Noether current is given by
\begin{equation}
J_{\text f}=\eta_\alpha \theta^\alpha- \tfrac1 2 T^\alpha_{\beta\gamma} \chi_\alpha \theta^\beta \theta^\gamma \ .\label{Noether}
\end{equation}
In computing $dJ_{\rm f}$ on-shell, the coefficient of the
$\chi^2 \theta^3$ term can be seen to vanish using
Eq. (\ref{G-GT-Tor}) and the Bianchi identity (\ref{BI-Rie})
for the Riemann tensor.
The coefficient of the $\eta\chi\theta^2$ term cancels by virtue of the relation between the curvatures of $\Gamma^\alpha_{\beta\gamma}$ and $\widetilde \Gamma^\alpha_{\beta\gamma}$ in (\ref{tildeR-R}) and the torsion Bianchi identity  (\ref{BI-Tor}).

\section{The gauged model}\label{sec:3}

\noindent In this section, following the standard Noether procedure adapted to the generalized Hamiltonian format, we shall derive a classical action for a version of the above sigma model in which the rigid supersymmetry is gauged.
That is, we shall add two fields to the model, namely a fermionic one-form
$\psi$, whose role is covariantize all the derivatives, and its fermionic zero-form Lagrange multiplier $\lambda$.
As we shall see, at the classical level, the gauging is consistent without imposing any further restrictions on the background Poisson geometry than those already underlying the construction of the ungauged model.

As already remarked, whether the model actually remains consistent at the quantum level, and whether there exists any related extension of Kontsevich's formality theorem to the deformation of differential Poisson algebras, remains to be investigated further.
In particular, it would be interesting to examine carefully potential anomalies in the Noether current and the conditions on the background in order to avoid or cancel these.
We would also like to remark that one may think of $\psi$ as a gravitino field for an abelian superalgebra without any bosonic generator,
which may be of importance in introducing further gaugings in order to construct topological open string models, that in particular could serve as microscopic
origins for higher spin gravities.

\subsection{The action}

\noindent In order to gauge the holonomic vector field $\delta_{\rm f}$ we introduce a fermionic one-form $\psi$ gauge field and its corresponding fermionic zero-form momentum $\lambda$, and couple $\psi$ to the Noether current \eqref{Noether}.
The resulting gauged model is described by the classical action
\be S=S_0+\int_{M_2} \left(\psi\wedge J_{\rm f}+ \lambda d\psi+\tfrac c 2 \psi\wedge \psi\right)\ ,\ee
or more explicitly,
\begin{eqnarray}
S&=&\int_{M_2} \left(
\eta_\a \wedge (d\phi^\a-\psi \theta^\a)+\tfrac12 \Pi^{\a\b} \eta_\a \wedge \eta_\b
+\chi_\a \wedge (d\theta^\a+(d\phi^\b-\psi\theta^\b)\Gamma^\a_{\b\gamma}\theta^\gamma) \right. \notag \\
&+&\left. \tfrac14 \widetilde R_{\gamma\d}{}^{\a\b} \chi_\a \wedge \chi_\b \theta^\gamma\theta^\d
+\lambda d\psi+\tfrac{c}2 \psi\wedge \psi\right)\ ,
\end{eqnarray}
where $c$ is a real parameter that cannot be fixed as we are gauging an abelian symmetry.
As for the degrees ${\rm deg}_2(\cdot)$ on $M_2$ and the additional Grassmann parity $\e_{\rm f}(\cdot)$, we have the following assignments:
\begin{center}
\begin{tabular}{| l |c | c | c | c | c | c |} \hline
 &  $\phi^\alpha$ & $\eta_ \alpha$ & $\theta^\alpha$ & $\chi^\alpha$ & $\psi$  & $\lambda$ \\ \hline
 $\text{deg}_2$ &  0 & 1 & 0 & 1 & 1 & 0 \\ \hline
 $\epsilon_{\text f}$ & 0 & 0 & 1 & 1& 1 & 1  \\ \hline
\end{tabular}
\end{center}

\subsection{Equations of motion}

We have the following equations of motion:
\begin{eqnarray}
{\cal R}^{\phi^\a} &:=& {\cal R}_0^{\phi^\a}-\psi \theta^\alpha=0\ ,\label{Rphi}\\[5pt]
{\cal R}^{\theta^\a} &:=& {\cal R}^{\theta^\a}_0 -\tfrac1 2 \psi\, T^\alpha_{\beta\gamma} \,\theta^\beta \theta^\gamma =0\ ,\label{Rtheta}\\[5pt]
{\cal R}^{\eta_\a} &:=& {\cal R}^{\eta_\a}_0 - \psi\wedge\left( R_{\a \gamma}{}^{\b}{}_{\d}\,\chi_\b\, \theta^\gamma \theta^\d  + T^\beta_{\alpha \gamma}  \,\eta_\beta\, \theta^\gamma+\tfrac 1 2 \widetilde R_{\beta \gamma}{}^\delta{}_\alpha  \,\chi_\delta \,\theta^\beta \theta^\gamma\right)   = 0  \label{Reta}\\
{\cal R}^{\chi_\a} &:=& {\cal R}^{\chi_\a}_0 +  \psi \wedge\left( \eta_\alpha+  T^\gamma_{\alpha \beta} \, \chi_\gamma\, \theta^\beta\right)= 0\ ,\label{Rchi} \\
\mathcal R^{\psi}&:=&d\psi=0\ , \label{Rpsi} \\
\mathcal R^{\lambda}&:=& d\lambda-J_{\rm f} -c \psi=0\ , \label{Rlambda}
\end{eqnarray}
The Cartan curvatures $({\cal R}^{\phi^\a}\,,
{\cal R}^{\theta^\a}\,,
{\cal R}^{\chi_\a}\,,\mathcal R^{\psi}\,,
\mathcal R^{\lambda})$ are proportional to the functional derivatives of $S$ with respect to ($\eta$, $\chi$, $\theta$, $\lambda$, $\psi$), respectively, while (\ref{Reta}) has been obtained from
\begin{eqnarray}
\frac{\delta S}{\delta \phi^\a}&=& d\eta_\alpha+\tfrac1 2 \partial_\alpha \Pi^{\beta\gamma} \eta_\beta \wedge \eta_\gamma+ \left(\Gamma^\gamma_{\alpha \beta}d\chi_\gamma -\chi_\gamma \wedge d\Gamma^\gamma_{\alpha \beta}+\partial_\alpha \Gamma^\gamma_{\delta \beta} \chi_\gamma \wedge (d\phi^\delta-\psi \theta^\delta) \right) \theta^\beta \notag \\[5pt]
&-&\Gamma^\gamma_{\alpha \beta}\chi_\gamma  \wedge d\theta^\beta+\tfrac 1 4  \partial_\alpha \widetilde R_{\beta \gamma}{}^{\delta\epsilon}\chi_\delta \wedge \chi_\epsilon \theta^\beta \theta^\gamma
\ ,
\end{eqnarray}
by rewriting $\partial_\alpha\Pi^{\beta\gamma} \eta_\beta \eta_\gamma$
using ${\cal R}^{\phi^\a}=0$ and $\widetilde \nabla_\alpha\Pi^{\b\gamma}=0$,
and the quantities $d\chi_\gamma \Gamma^\gamma_{\alpha \beta} \theta^\beta$
and $\chi_\gamma \Gamma^\gamma_{\alpha\beta} d\theta^\beta$
using ${\cal R}^{\chi_\a}= 0$ and ${\cal R}^{\theta^\a}=0$, respectively.

Alternatively, upon defining
\begin{eqnarray}
D\phi^\a&=&d\phi^\a-\psi\theta^\a \ , \qquad\qquad\qquad\quad
D\theta^\a=\nabla\theta^\a -\tfrac12 \psi\theta^\b T_{\b\gamma}^\a \theta^\gamma \ , \\
D\eta_\a&=&\nabla\eta_\a+\tfrac12 \psi\theta^\b T_{\b\a}^\gamma \wedge \eta_\gamma\ , \qquad
D\chi_\a=\nabla\chi_\a+\tfrac12 \psi\theta^\b T_{\b\a}^\gamma \wedge \chi_\gamma\ ,
\end{eqnarray}
the equations of motion take the form
\bea&& D\phi^\a+\Pi^{\a\b}\eta_\b=0\ ,\\[5pt]
&& D\theta^\a+\tfrac12 \widetilde R_{\gamma\d}{}^{\a\b}\,\chi_\b\,\theta^\gamma\theta^\d=0\ ,\\[5pt]
&& D\chi_\a-\tfrac12 \widetilde R_{\a\b}{}^{\gamma\d}\chi_\gamma \wedge \chi_\d\theta^\b+\psi\wedge\left(\eta_\a+
\tfrac12 T^\gamma_{\a\b} \, \chi_\gamma\, \theta^\b\right)=0\ ,\\[5pt]
&&D\eta_\a+R_{\a\gamma}{}^\b{}_\d\,\chi_\b \wedge D\phi^\gamma \theta^\d+
\tfrac 1 4  \nabla_\a \widetilde R_{\d\e}{}^{\b\gamma}\chi_\b \wedge \chi_\gamma \, \theta^\d\theta^\e -\tfrac 1 2 \psi \wedge\left(\widetilde R_{\beta \gamma}{}^\delta{}_\alpha \, \chi_\delta \,\theta^\beta \theta^\gamma+ T_{\a\b}^\gamma \, \eta_{\gamma} \theta^\b\right) = 0\ .\qquad\eea
This more geometric form of writing the equations is less useful, however,
for verifying the Cartain integrability, as the on-shell value of the
$D^2$ operator is more involved than that of the $\nabla^2$ operator, as
can be seen by comparing
\begin{eqnarray}
D^2 \phi^\a&=& \tfrac12 \Pi^{\b\delta} \Pi^{\gamma\epsilon} T^\a_{\b\gamma} \eta_{\delta}\wedge\eta_{\epsilon}-\tfrac12 \psi \wedge\left( \Pi^{\gamma\delta} T^\a_{\b\gamma} \eta_{\delta} \theta^\b+\widetilde R_{\gamma\delta}{}^{\a\b}  \chi_\b \theta^\gamma
\theta^\delta\right)\ ,\\[5pt]
 D^2\theta^\a&=&-\psi \wedge\left( \Pi^{\gamma\e} R_{\b\gamma}{}^\a{}_\d  \eta_\e  \theta^\b \theta^\d
+\tfrac14  T^\a_{\b\gamma}  \widetilde R_{\d\e}{}^{\b\r}  \chi_\r\theta^\d\theta^\e
 \theta^\gamma +\tfrac12   \Pi^{\d\r}
\nabla_\d T^\a_{\b\gamma}  \eta_\r \theta^\b \theta^\gamma\right)\ ,\qquad \\[5pt]
D^2\chi_\a &=&- \psi\wedge\left( \Pi^{\d\e}  R_{\gamma\d}{}^\b{}_\a  \eta_\e \wedge \chi_\b \theta^\gamma
-\tfrac14  T^\gamma_{\b\a} \widetilde R_{\d\e}{}^{\b\r}  \chi_\r \wedge \chi_\gamma \theta^\d\theta^\e
+\tfrac12   \Pi^{\d\e}  \nabla_\d T^\gamma_{\b\a} \eta_\e \wedge \chi_\gamma \theta^\b  \right)\notag \\[5pt]
&&+\tfrac12 \psi \wedge \psi \wedge\left( R_{\gamma\d}{}^\b{}_\a  \chi_\b \theta^\gamma\theta^\d
-\tfrac12  T^\b_{\d\r} T^\gamma_{\b\a}  \chi_\gamma \theta^\d \theta^\r
+   \nabla_\d T^\gamma_{\b\a}  \chi_\gamma \theta^\b\theta^\d  +\tfrac12  T^\gamma_{\b\a} T^\r_{\e\gamma}\chi_\r  \theta^\b\theta^\e \right) \ ,\qquad
\end{eqnarray}
and a similar expression for $D^2\eta_\a$,
with
\begin{eqnarray}
\nabla d\phi^\a&=& \tfrac12 \Pi^{\b\delta} \Pi^{\gamma\epsilon} T^\a_{\b\gamma} \eta_{\delta}\wedge\eta_{\epsilon}-\psi \wedge \Pi^{\gamma\d} T^\a_{\b\gamma} \,\eta_\d\, \theta^\b -
\tfrac12 \psi\wedge\psi\, T^\a_{\b\gamma}\th^\b \th^\gamma\ ,\label{dd1}\\[5pt]
\nabla^2\theta^\a&=&-\psi\wedge \Pi^{\d\e} R_{\gamma\d}{}^\a{}_\b \eta_\e \theta^\gamma\theta^\b -\tfrac12 \psi\wedge\psi R_{\gamma\d}{}^\a{}_\b \th^\gamma\th^\d \th^\b\ ,\qquad \label{dd2}\\[5pt]
\nabla^2\eta_\a &=& \psi\wedge \Pi^{\d\e} R_{\gamma\d}{}^\b{}_\a \eta_\e \wedge \eta_\b\theta^\gamma +\tfrac12 \psi\wedge\psi R_{\gamma\d}{}^\b{}_\a \eta_\b \th^\gamma\th^\d\ , \label{dd3}\\[5pt]
\nabla^2\chi_\a &=& -\psi\wedge \Pi^{\d\e} R_{\gamma\d}{}^\b{}_\a \eta_\e\wedge \chi_\b\theta^\gamma +\tfrac12 \psi\wedge\psi R_{\gamma\d}{}^\b{}_\a \chi_\b \th^\gamma\th^\d\ ,\label{dd4}
\end{eqnarray}
where $ \Pi^{\a\b} \Pi^{\c\d} R_{\b\d}{}^\e{}_\lambda=0$ has been used
repeatedly, and the torsion Bianchi identity Eq. \eqref{BI-Tor} has been
used to cancel the $\psi^2$-terms in $D^2\theta^\a$.
Thus, in what follows, we shall verify the integrability of the equations of
motion by acting with $\nabla$ on the generalized Cartan curvatures.

\subsection{Universal Cartan integrability}
\noindent Letting ${\cal R}^i=({\cal R}^{\phi^\a},{\cal R}^{\theta^\a},
{\cal R}^{\eta_\a},{\cal R}^{\chi_\a}, {\cal R}^{\psi}, {\cal R}^{\lambda} )$, sometimes referred to as the Cartan curvatures, we are going to show that there exists
a field dependent matrix $M^i_j$ such that the obstructions
\be {\cal A}^i:=\nabla {\cal R}^i +  {\cal R}^j \wedge M^i_j \label{obs}\ee
defined off-shell, vanish on base manifolds of arbitrary dimensions, essentially by virtue of Eqs. \eqref{J000}--\eqref{J111} and the compatibility condition
\eqref{comp1} and its consequences \eqref{comp2} and \eqref{Pi(R-R-NT)}.
In other words, the equations of motion form a universally Cartan integrable system on the base manifold, that is, the field variables generate a free differential algebra on-shell, by virtue of the fact that the target space geometry describes a differential Poisson algebra.

In order to analyze the above problem, one first computes
$\nabla {\cal R}^i$ off-shell, and one may then proceed on-shell
by first using the identities \eqref{dd1}--\eqref{dd4}
and then simplify the expressions further by means of
identities involving the background fields.
In doing so, we shall not keep track of the matrix $M^i_j$,
and we shall expand ${\cal A}^i={\cal A}^i_0+\psi\wedge {\cal A}^i_1+\tfrac12 \psi\wedge\psi \wedge {\cal A}^i_2$, where thus ${\cal A}^i_0$ vanish due to Eqs. \eqref{J000}--\eqref{J111}, \eqref{comp1} and \eqref{comp2} (but not \eqref{Pi(R-R-NT)}), as was shown in \cite{us}.
Thus, we only need to analyze the potential
obstructions from ${\cal A}^i_1$ and ${\cal A}^i_2$.
As we have already stressed, we shall see that these obstructions vanish
without any further conditions on the background (though the vanishing of ${\cal A}^{\eta_\a}$ requires Eq. \eqref{Pi(R-R-NT)}).\\[-10pt]

\noindent{${\cal A}^{\psi}$:} This quantity vanishes identically because $\nabla d\psi=dd\psi=0$.\\[-10pt]

\noindent{${\cal A}^{\lambda}$:} As $\nabla d\lambda=0$ and $\nabla \psi=d\psi=0$, we have to check that $dJ_{\text f}$ remains closed on-shell in the gauged model.
Indeed, the $\psi \chi\theta^3$ term vanishes identically due to the relation (\ref{tildeR-R}) and the torsion Bianchi identity.\\[-10pt]

\noindent{${\cal A}^{\phi^\a}$:} The $\psi \chi \theta^2$ and $\psi^2 \theta^2$ terms cancel uneventfully while the coefficient of the $\psi \eta \theta$ term cancel as a consequence of the compatibility assumption (\ref{NTPi}).\\[-10pt]

\noindent{${\cal A}^{\theta^\a}$:} The $\psi\eta \theta^2$ term can be seen to cancel using the relation (\ref{tildeR-R}) and the torsion Bianchi identity (\ref{BI-Tor}).
The $\psi\chi \theta^3 $ term vanishes by virtue of (\ref{G-GT-Tor}) and the Bianchi identity (\ref{BI-Rie}) for the Riemann tensor of $\widetilde \Gamma$.
The $\psi^2 \theta^3$ term can be cancelled using the Bianchi identity for the Riemann tensor of $\Gamma$.\\[-10pt]

\noindent{${\cal A}^{\chi_\a}$:} The $\psi\eta\chi\theta$ term can be cancelled by first using (\ref{tildeR-R}) and then the torsion Bianchi identity (\ref{BI-Tor}).
The $\psi\chi^2\theta^2$ terms caan be seen to vanish by first using (\ref{G-GT-Tor})
and then the Bianchi identity (\ref{BI-Rie}) for the Riemann tensor of the connection $\tilde \Gamma$.
Finally, the $\psi^2\theta^2\chi$ terms cancel due to (\ref{tildeR-R}).\\[-10pt]

\noindent{${\cal A}^{\eta_\a}$:} The verification of the vanishing of this obstruction is less straightforward than the previous ones.
The $\psi \eta^2 \theta$ term cancels due to the identity (\ref{Pi(R-R-NT)}) (which was
not needed in checking the Cartan integrability of the ungauged model).
The coefficient of the $\psi \chi^2 \theta^3$ term is proportional to the covariant derivative of the Bianchi identity for the Riemann tensor of the connection $\tilde \Gamma$, that is, $\nabla_\mu(\widetilde \nabla_{[\a} \widetilde R_{\b\gamma]}{}^{\d\e}- \widetilde T^\lambda_{[\a\b} \widetilde R_{\gamma] \lambda}{}^{\d_\e})=0$.
One can furthermore check that the $\psi \chi \eta \theta^2$ term vanishes using (\ref{tildeR-R}) and the two Bianchi identities (\ref{BI-Rie}) and (\ref{BI-Tor}).
The coefficient of the $\psi^2 \eta \theta^2$ term can be seen to vanish using (\ref{tildeR-R}).
Finally, the $\psi^2 \chi \theta^3$ term can be cancelled by first using (\ref{G-GT-Tor}) and then the Bianchi identity (\ref{BI-Rie}) for the Riemann tensor.

\subsection{Gauge transformations}

\noindent Following the general procedure \cite{Boulanger2012}
outlined in \cite{us}, we use $\mathcal R^{\phi^\a}$
to eliminate $d\phi^\a$,
as to write the equations of motion as
\begin{eqnarray}
\widehat {\cal R}^{\phi^\a}&=&d\phi^\a+ \Pi^{\a\b}\eta_\b - \psi \theta^\alpha\ ,\\[5pt]
\widehat {\cal R}^{\theta^\a}&=&d \theta^\a-\Pi^{\beta \gamma} \Gamma^\alpha_{\beta \delta}  \eta_\gamma \theta^\delta + \tfrac 1 2 \widetilde R_{\gamma\d}{}^{\a\b} \chi_\b \theta^\gamma \theta^\d \ ,\eea\\[-35pt]
\bea
\widehat {\cal R}^{\eta_\a}&=&d \eta_\a + \Pi^{\beta \gamma} \Gamma^\delta_{\beta \alpha}  \,\eta_\gamma \wedge \eta_\delta  +\Pi^{\gamma\lambda} R_{\a\gamma}{}^\b{}_\d \,\eta_\lambda
\wedge\chi_\b \theta^\delta+\tfrac{1}4  \nabla_\a \widetilde R_{\d\e}{}^{\b\gamma} \chi_\b \wedge \chi_\gamma \theta^\d\theta^\e \ , \qquad\nn\\[5pt]
 &-&\psi \wedge \left( \Gamma^\beta_{\alpha \gamma} \eta_\beta \theta^\gamma  +\tfrac 1 2\widetilde R_{\beta \gamma}{}^\delta{}_\alpha \chi_\delta \theta^\beta \theta^\gamma \right) \ , \eea\\[-35pt]
\be \widehat {\cal R}^{\chi_\a}=d \chi_\a +
\Pi^{\beta \gamma} \Gamma^\delta_{\beta \alpha}  \eta_\gamma \wedge \chi_\delta - \tfrac 1 2  \widetilde R_{\a\d}{}^{\b\gamma} \chi_\b  \wedge \chi_\gamma \theta^\d+ \psi \wedge \left( \eta_\alpha+ \Gamma^\beta_{\alpha \gamma} \chi_\beta \theta^\gamma \right) \ ,\ee\\[-35pt]
\bea
\widehat {\mathcal R}^{\psi}&=&d\psi=0\ , \\[5pt]
\widehat {\mathcal R}^{\lambda}&=& d\lambda-\eta_\a \theta^\a+\tfrac12 T^\a_{\b\gamma} \chi_\a
\theta^\b\theta^\gamma-c \psi=0\ .
\end{eqnarray}
Since these curvatures are on the canonical form
\begin{equation}
\widehat{\mathcal R}^i := dZ^i+ \widehat {\mathcal Q}^i(Z^j) = 0\ ,\qquad
Z^i=(\phi^\a,\eta_\a;\theta^\a,\chi_\a; \psi, \lambda)\ ,
\end{equation}
where thus $\widehat {\mathcal Q}^i$ is a Q structure, \emph{viz.}
\be \widehat {\mathcal Q}^i\wedge \frac{\partial}{\partial Z^j} \widehat {\mathcal Q}^j=0\ ,\ee
the universal Cartan integrability amounts to the generalized Bianchi identities
\be d\widehat {\mathcal R}^i+\widehat {\mathcal R}^j\wedge \frac{\partial}{\partial Z^j} \widehat {\mathcal Q}^i= 0\ .\ee
The on-shell gauge transformations are consequently given by
\begin{equation}
\delta Z^i=d \epsilon^i - \epsilon^j \frac{\partial}{\partial Z^j} \widehat {\mathcal Q}^i\ ,  \qquad
\mbox{modulo $\widehat {\cal R}^i$}\ ,
\end{equation}
where $\epsilon^i=( 0, \epsilon^{(\eta)}_ \alpha;0, \epsilon^{(\chi)}_{\alpha}; \epsilon^{(\psi)},0  )$ are gauge parameters with degrees and Grassmann parities
given by
\begin{center}
\begin{tabular}{| l |c | c | c |} \hline
 &  $\epsilon^{(\eta)}_ \alpha$ & $\epsilon^{(\chi)}_ \alpha$ & $\epsilon^{(\psi)}$ \\ \hline
 $\text{deg}_2$ &  0 & 0 & 0 \\ \hline
 $\epsilon_{\text f}$ & 0 & 1 & 1 \\ \hline
\end{tabular}
\end{center}
Applying the above general formalism to the present model, the infinitesimal on-shell gauge transformations are found to be
\begin{eqnarray}
\delta \phi^\alpha&=&-\Pi^{\alpha \beta} \epsilon^{(\eta)}_{\beta}+\epsilon^{(\psi)} \theta^\alpha \ ,\\[5pt]
\delta \theta^\alpha&=&   \Pi^{\beta \gamma} \Gamma^\alpha_{\beta \delta} \epsilon^{(\eta)}_{ \gamma} \theta^\delta  -\tfrac 1 2  \widetilde R_{\gamma \delta}{}^{\alpha \beta} \epsilon^{(\chi)}_{ \beta} \theta^\gamma \theta^\delta\ ,\\[5pt]
\delta \eta_\alpha &=& \nabla \epsilon^{(\eta)}_{\alpha}- \Pi^{\beta \gamma} \Gamma^{\delta}_{\beta \alpha}  \epsilon^{(\eta)}_\gamma \eta_\delta  -\Pi^{\gamma \lambda} R_{\a\gamma}{}^\b{}_\d \,\epsilon^{(\eta)}_\lambda \chi_\b \theta^\delta
+\Pi^{\gamma \lambda} R_{\a\gamma}{}^\b{}_\d \,\eta_\lambda
\epsilon^{(\chi)}_\b \theta^\delta\nn\\[5pt]&&-\tfrac 1 2  \nabla_\a \widetilde R_{\d\e}{}^{\b\gamma} \epsilon^{(\chi)}_\b \chi_\gamma \theta^\d\theta^\e\ + \epsilon^{(\psi)}  \left( \Gamma^\beta_{\alpha \gamma} \eta_\beta \theta^\gamma  +\tfrac 1 2\widetilde R_{\beta \gamma}{}^\delta{}_\alpha \chi_\delta \theta^\beta \theta^\gamma \right),\\[5pt]
\delta \chi_\alpha  &=& \nabla \epsilon^{(\chi)}_\alpha-  \Pi^{\beta \gamma} \Gamma^\delta_{\beta \alpha} \epsilon^{(\eta) }_{\gamma}\chi_\delta+ \widetilde R_{\alpha \delta}{}^{\beta \gamma} \epsilon^{(\chi) }_{\beta} \chi_\gamma \theta^\delta- \epsilon^{(\psi)} \left( \eta_\alpha+ \Gamma^\beta_{\alpha \gamma} \chi_\beta \theta^\gamma \right) \ , \\[5pt]
\delta \psi&=&d \epsilon^{(\psi)} \ , \\[5pt]
\delta \lambda&=&\epsilon^{(\eta)}_\alpha \theta^\a-\tfrac12 T^\a_{\b\gamma} \epsilon^{(\chi)}_\a
\theta^\b\theta^\gamma +c \epsilon^{(\psi)}\ .
\end{eqnarray}
In particular, we note that the gauge transformation
associated to the gauge parameter $\epsilon^{(\psi)}$
corresponds to the $\delta_\text f$ supersymmetric
transformation (\ref{susy}), \emph{i.e.}
\be
\delta_{\epsilon^{(\psi)}} (\phi^\a, \eta_\a; \theta^\a,\chi_\a)=\epsilon^{(\psi)} \delta_{\rm f} (\phi^\a, \eta_\a; \theta^\a,\chi_\a)\ ,
\ee
and moreover
\be
\delta_{\epsilon^{(\psi)}} \psi=d \epsilon^{(\psi)} \ ,\qquad
\delta_{\epsilon^{(\psi)}} \lambda= c\, \epsilon^{(\psi)} \ .\ee
Off-shell, it follows from the general formalism
\cite{Boulanger2012}, that the gauge transformations
are given by
\be \delta Z^i=d \epsilon^i - \epsilon^j \frac{\partial}{\partial Z^j} \widehat{{\mathcal Q}}^i+
\tfrac12 \epsilon^k \,\widehat{\cal R}^l \,\frac{\partial \Omega_{kj}}{\partial Z^l} \,{\cal P}^{ji}\ ,\label{nic}\ee
where we have introduced the symplectic two-form
\be {\Omega}=d \vartheta =\tfrac12 dZ^i {\cal O}_{ij} dZ^j=\tfrac12 dZ^i dZ^j \Omega_{ij}\ ,\qquad  {\cal P}^{ik}{\cal O}_{kj}=-\delta^i_j\ ,\ee
of degree three on the full target space ${\cal N}$, with pre-symplectic structure
\be \vartheta =\eta_\a \wedge d\phi^\a+\chi_\a \wedge \nabla \theta^\a + \lambda d \psi \ ,\ee
treated as a one-form of $\mathbb N$-degree two on ${\cal N}$.
Thus, the matrix $ {\cal O}_{ij}$ can be read off from
\begin{equation}
 \Omega= \tfrac1 2\begin{pmatrix}
d\phi^\rho & d\eta_\rho & d\theta^\rho & d\chi_\rho & d\psi & d\lambda
\end{pmatrix}
\begin{pmatrix}
2 \partial_{[\rho} \Gamma^\alpha_{\gamma]\beta} \chi_\alpha \theta^\beta \quad &  \delta_\rho{}^\gamma \quad& -\Gamma^\alpha_{\rho \gamma} \chi_\alpha \quad& -\Gamma^\gamma_{\rho \alpha} \theta^\alpha & 0 & 0 \\
\delta^\rho{}_\gamma &  0 & 0 & 0 & 0 & 0\\
-\Gamma^\alpha_{\gamma \rho} \chi_\alpha &  0 & 0 & -\delta_\rho{}^\gamma& 0 & 0  \\
\Gamma^\rho_{\gamma\alpha} \theta^\alpha &  0 & \delta^\rho{}_\gamma & 0 & 0 & 0 \\
0 & 0 & 0 & 0 & 0 & -1 \\
0 & 0 & 0 & 0 & 1 & 0
\end{pmatrix}
\begin{pmatrix}
d\phi^\gamma \\ d\eta_\gamma \\ d\theta^\gamma \\ d\chi_\gamma \\ d\psi \\ d\lambda
\end{pmatrix}\ .
\end{equation}
Moreover, the components ${\cal P}^{ji}$ of the Poisson structure on ${\cal N}$ is given by
\begin{equation}
{\cal P}^{ik}=\begin{pmatrix}
0 & - \delta^\sigma{}_\rho & 0 & 0 & 0 & 0 \\
-\delta_\sigma{}^\rho \quad &  R_{\sigma\rho}{}^\alpha{}_\beta \chi_\alpha \theta^\beta \quad& \Gamma^\rho_{\sigma\alpha}\theta^\alpha \quad & -\Gamma^\alpha_{\sigma\rho} \chi_\alpha  & 0 & 0 \\
0 &  \Gamma^\sigma_{\rho\alpha} \theta^\alpha & 0 & -\delta^\sigma{}_\rho & 0 &  0\\
0 &  \Gamma^\alpha_{\rho\sigma} \chi_\alpha & \delta_\sigma{}^\rho & 0 & 0 & 0 \\
0 & 0 & 0 & 0 & 0 & -1 \\
0 & 0 & 0 & 0 & 1 & 0
\end{pmatrix} \ .
\end{equation}
Using the above four by four matrices is simple to show that ${\cal P}^{ik}{\cal O}_{kj}=-\delta^i_j$.
If the connection vanishes identically, then the off-shell modification
of the gauge transformation \eqref{nic} vanishes.

\section{Conclusions and remarks}\label{sec:4}

\noindent Starting from the two-dimensional topological sigma model that was constructed in \cite{us} with the aim of covariantly quantizing differential Poisson algebras, we  have shown that it is possible to gauge its rigid supersymmetry without having to impose any further constraints on the background fields.
It remains to be examined whether the gauging is consistent at the
quantum level, which in particular would require that the charge
\be q_{\rm f}=\oint J_{\rm f}\ ,\ee
of the Noether current \eqref{Noether}, or a suitably improved version
thereof, is nilpotent as an operator, which may lead to extra
conditions on the geometric structures of the differential Poisson
algebra, \emph{e.g.} its Ricci curvature $R_{\a\b}{}^\b{}_{\gamma}$.
However, if such obstructions were to arise, one might argue that they
would do so already in the star product at the first sub-leading order
in the deformation parameter, but this is not the case, as was shown
in \cite{Zumino} following the algebraic approach to deformation quantization.
This suggests that there will not arise any obstruction at any
order of perturbation theory, which we plan to examine in a
separate work.

One may ask for several natural generalizations of the present model:
First of all, it is possible to rewrite the action on a format more closely
related to the original Ikeda--Schaller--Strobl Poisson model \cite{Cesar}
which facilitates the inclusion of the tensorial one-form
$S^{\a\b}_\gamma$ in Eq. \eqref{S} while preserving the rigid supersymmetry,
as well as further gaugings resulting in more general $Q$-structures
than the de Rham differential.
For example, it would be interesting to gauge additional
holonomic vector fields arising from complex structures
on complex submanifolds, or $sp(2)$ algebras associated to
submanifolds with conical metric structures.

Assuming quantum consistency, it is natural to expect that
finite deformations of the background can be modeled by
an open string field $\Psi$ obeying
\be \{Q,\Psi\}_\star + \Psi\star \Psi=0\ ,\label{QPsi}\ee
using the BRST operator
\be Q = \oint \gamma q_{\rm f}+ Q^{(\eta,\chi)}\ ,\ee
where $\gamma$ is a bosonic ghost $\gamma$ and $Q^{(\eta,\chi)}$
is built from the fermionic $\eta$ and $\chi$ ghosts.
Drawing on recent results in higher spin gravity \cite{FCSDD, FCS4D},
we propose that $\Psi$ contains a superconnection
$Z$ valued in the direct product of two copies of
the differential star product algebra on $N$, arising
from zero-modes and winding-modes, and finite-dimensional
matrices forming a graded Frobenius algebra ${\cal F}$ with
inner Klein operator $\kappa$.
Saturating the fermionic zero-modes by inserting
a delta function in the path integral, one may then argue
that \eqref{QPsi} contains the flatness condition
\be \{\kappa q_{\rm f},Z\}_\star +Z\star Z=0\ ,\ee
where $q_{\rm f}$ thus acts only on winding modes,
and $Z$ consists of even forms in the odd part of
${\cal F}$ and odd forms in the even part of ${\cal F}$,
all valued in the associative (higher spin) algebra generated
by the bosonic zero-modes.
Thus, if confirmed, our proposal would provide a first-quantized framework
for the duality extended Frobenius--Chern--Simons
formulation of higher spin gravity given in \cite{FCSDD,FCS4D}.

\paragraph{Acknowledgements:} We would like thank Cesar Arias,
Nicolas Boulanger and Andrew Waldron for discussions.
R. B. would like to thank the hospitality of UNAB during the completion of this work.
A. T. G. is supported by Fondecyt post-doc grant N$^{\rm o}$ 3130333.
The work of P.S. is supported by Fondecyt Regular grant N$^{\rm o}$ 1140296 and Conicyt grant DPI 20140115.

\begin{appendix}
\section{Conventions and some useful identities}\label{app:a}

\noindent The covariant exterior derivative of a one-form $\omega=\omega_\alpha d\phi^\alpha$ is given by
\be
\nabla \omega_{\a} = d\omega_{\a} - \C^\b{}_\a \wedge \omega_{\b}\ ,\qquad
\Gamma^\a{}_\b=d\phi^\gamma \Gamma^{\a}_{\gamma \b}\ .
\ee
In terms of components, we have
\be
\nabla \omega_\a=d\phi^\beta \nabla_\b \omega_\a\ ,\qquad
\nabla_{\a} \o_{\b} = \partial_{\a} \o_{\b} - \C_{\a \b}^{\gamma} \o_{\gamma}\ .
\ee
The curvature and torsion tensors
\be
 R_{\a\b}{}^{\gamma}{}_{\d}= 2\, \partial_{[\a} \C_{\b]\d}^{\gamma} + 2\, \C_{[\a|\e}^{\gamma} \C_{|\b]\d}^{\e}  \ ,\qquad
 T_{\a\b}^{\gamma} = 2\, \C_{[\a\b]}^{\gamma}\ .
\ee
The square of the exterior covariant derivative acting on a one-form is
\be  \nabla^2 \omega_\a= - R^\b{}_\a \wedge \omega_\b\ ,\qquad R^\b{}_\a = \tfrac12 d\phi^\c\wedge d\phi^\d R_{\c\d}{}^\b{}_\a\ .\ee
We also make use of the covariant derivative $\widetilde \nabla$ with connection one-form %
\begin{equation}\label{G-GT-Tor}
\widetilde \Gamma^\alpha{}_{\beta}=d\phi^\gamma \widetilde\Gamma^\a_{\gamma\beta}=
d\phi^\c\Gamma^\a_{\b\c}\ ,\qquad \widetilde \Gamma^\alpha_{\beta\gamma}=\Gamma^\a_{\c\b}=
\Gamma^\alpha_{\beta\gamma}-T^\alpha_{\beta\gamma}\ ,
\end{equation}
and curvature two-form
\be \widetilde\nabla^2 \omega_\a=-\widetilde R^\b{}_\a \wedge \omega_\b\ ,\qquad
\widetilde R^\a{}_\b=\tfrac12 d\phi^\c\wedge d\phi^\d \widetilde R_{\c\d,}{}^\a{}_\b\ ,\ee
where
\begin{equation}\label{tildeR-R}
\widetilde R_{\alpha\beta}{}^\gamma{}_\delta=R_{\alpha\beta}{}^\gamma{}_\delta-2 \nabla_{[\alpha} T^\gamma_{\beta]\delta}+2 T^\gamma_{[\alpha| \epsilon}  T^\epsilon_{| \beta] \delta} - T^\epsilon_{\alpha\beta}  T^\gamma_{\epsilon \delta} \ .
\end{equation}
The component forms of the Bianchi identities for the Riemann tensor and the torsion read as follows:
\begin{equation}\label{BI-Rie}
\nabla_{[\a} R_{\b\gamma]}{}^{\d}{}_\e- T^\lambda_{[\a\b} R_{\gamma] \lambda}{}^\d{}_\e=0\ ,
\end{equation}
and
\begin{equation}\label{BI-Tor}
R_{[\a\b}{}^{\gamma}{}_{\d]} =\nabla_{[\a} T_{\b\d]}^{\gamma} - T_{[\a\b}^{\e} T_{\d] \e}^{\gamma} \ ,  \\
\end{equation}
and, similarly, for the $\widetilde \Gamma$ connection we have $\widetilde \nabla_{[\a} \widetilde R_{\b\gamma]}{}^{\d}{}_\e- \widetilde T^\lambda_{[\a\b} \widetilde R_{\gamma] \lambda}{}^\d{}_\e=0$ and $\widetilde R_{[\a\b}{}^{\gamma}{}_{\d]} =\widetilde \nabla_{[\a} \widetilde T_{\b\d]}^{\gamma} - \widetilde T_{[\a\b}^{\e} \widetilde T_{\d] \e}^{\gamma}$.

\noindent The compatibility condition $ \widetilde \nabla_\alpha \Pi^{\beta \gamma}=0$ can equivalently be written as
\be \label{NTPi}
\nabla_\a \Pi^{\b\gamma}= -2 \Pi^{\delta[\beta}T^{\gamma]}_{\a\d} \ .\ee
The integrability of the compatibility condition implies that
\be
\widetilde R_{\alpha \beta}{}^{\gamma\delta}:= \Pi^{\delta \epsilon} \widetilde R_{\alpha\beta}{}^{\gamma}{}_\epsilon = \widetilde R_{\alpha \beta}{}^{\d\c}\ .
\ee
Alternatively, starting from the Ricci identity applied to the Poisson bi-vector, \emph{viz.}
 \begin{equation}
[\nabla_\epsilon, \nabla_\alpha ] \Pi^{\beta\gamma}= -T^\delta_{\epsilon \alpha} \nabla_\delta \Pi^{\beta\gamma}-2 R_{\epsilon\alpha}{}^{[\beta}{}_\delta \Pi^{\gamma]\delta}\ ,
\end{equation}
and using the identity (\ref{BI-Tor}), we find that the integrability of the  compatibility condition (\ref{NTPi}) can be phrased as that
\begin{equation}
\Pi^{\lambda [\beta}\left( R_{\lambda \epsilon}{}^{\gamma]}{}_\alpha+ R_{\alpha \lambda}{}^{\gamma]}{}_\epsilon- \nabla_\lambda T^{\gamma]}_{\epsilon \alpha} \right) =0 \ .
\end{equation}

\end{appendix}



\begin{thebibliography}{99}

\bibitem{Ikeda}
N. Ikeda, ``\textit{Two-dimensional gravity and nonlinear gauge theory}", \href{http://dx.doi.org/10.1006/aphy.1994.1104}{Annals Phys. 235 (1994) 435-464}, \href{http://arxiv.org/abs/hep-th/9312059}{arXiv:hep-th/9312059}.

\bibitem{Schaller}
P. Schaller, T. Strobl, ``\textit{Poisson structure induced (topological) field theories}, \href{http://dx.doi.org/10.1142/S0217732394002951}{Mod.Phys.Lett. A9 (1994) 3129-3136}, \href{http://arxiv.org/abs/hep-th/9405110}{arXiv:hep-th/9405110}.

\bibitem{CaFe}
A. S. Cattaneo and G. Felder, ``\textit{A Path integral approach to the Kontsevich quantization formula}", \href{http://dx.doi.org/10.1007/s002200000229}{Commun.Math.Phys. 212 (2000) 591-611}, \href{http://arxiv.org/abs/math/9902090}{arXiv:math/9902090}.

\bibitem{Kontsevich}
M. Kontsevich, ``\textit{Deformation quantization of Poisson manifolds}", \href{http://dx.doi.org/10.1023/B:MATH.0000027508.00421.bf}{Lett.Math.Phys. 66 (2003) 157-216 }, \href{http://arxiv.org/abs/arXiv:q-alg/9709040}{arXiv:q-alg/9709040 [q-alg] }.

\bibitem{Chu}
C.~Chu and P.~ Ho,``\textit{Poisson Algebra Of Differential Forms}'', \href{http://dx.doi.org/10.1142/S0217751X97002929}{Int.J.Mod.Phys. 12 (1997) 5573-5587},
\href{http://arxiv.org/abs/q-alg/9612031}{arXiv:q-alg/9612031}.

\bibitem{Beggs}
E. J. Beggs and S. Majid, ``\textit{Semiclassical differential structures}", \href{http://arxiv.org/abs/math/0306273}{arXiv:math/0306273}.

\bibitem{Tagliaferro}
  A.~Tagliaferro,
  ``\textit{A Star Product for Differential Forms on Symplectic Manifolds}'',
  \href{http://arxiv.org/abs/0809.4717v2}{arXiv:0809.4717 [hep-th]}.

\bibitem{Zumino}
  S.~McCurdy and B.~Zumino,
  ``\textit{Covariant Star Product for Exterior Differential Forms on Symplectic Manifolds}'', \href{http://dx.doi.org/10.1063/1.3327559}{AIP Conf.Proc. 1200 (2010) 204-214},
  \href{http://arxiv.org/abs/0910.0459}{arXiv:0910.0459 [hep-th]}.

\bibitem{Finnish}
M. Chaichian, M. Oksanen, A. Tureanu and G. Zet, ``\textit{Covariant star product on symplectic and Poisson spacetime manifolds}, \href{http://dx.doi.org/10.1142/S0217751X10049785}{Int.J.Mod.Phys. A25 (2010) 3765-3796}, \href{http://arxiv.org/abs/arXiv:1001.0503}{arXiv:1001.0503 [math-ph]}.

\bibitem{us}
C. Arias, N. Boulanger, P. Sundell and A. Torres-Gomez, ``\textit{2D sigma models and differential Poisson algebras}", \href{http://arxiv.org/abs/arXiv:1503.05625}{arXiv:1503.05625 [hep-th]}.



\bibitem{GR_as_CS}
R. Bonezzi, O. Corradini and A. Waldrom, ``\textit{Is Quantum Gravity a Chern-Simons Theory? }", \href{http://dx.doi.org/10.1103/PhysRevD.90.084018}{Phys.Rev. D90 (2014) 8, 084018 }, \href{http://arxiv.org/abs/arXiv:1407.5977}{arXiv:1407.5977 [hep-th]}.


\bibitem{FCSDD}
N. Boulanger, E. Sezgin and P. Sundell, ``\textit{4D higher spin gravity with dynamical two-form as a Frobenius--Chern--Simons gauge theory}", to appear.

\bibitem{FCS4D}
R. Bonezzi and P. Sundell, ``\textit{Higher Spin Gravity in any Dimension with Dynamical Two-form}", to appear


\bibitem{Park}
J. S. Park, ``\textit{Topological open p-branes}'',
\href{http://arxiv.org/abs/hep-th/0012141}{arXiv:hep-th/0012141}.

\bibitem{Ikeda2001}
N. Ikeda, ``\textit{Deformation of BF theories, topological open membrane and a generalization of the star deformation}, \href{http://dx.doi.org/10.1088/1126-6708/2001/07/037  }{JHEP 0107 (2001) 037}, \href{http://arxiv.org/abs/hep-th/0105286}{arXiv:hep-th/0105286}.

\bibitem{Hofman1}
C. Hofman and J. S. Park, ``\textit{Topological open membranes}", \href{http://arxiv.org/abs/hep-th/0209148}{arXiv:hep-th/0209148}.

\bibitem{Hofman2}
C. Hofman and J. S. Park, ``\textit{BV quantization topological open membranes}, \href{http://dx.doi.org/10.1007/s00220-004-1106-7}{Commun.Math.Phys. 249 (2004) 249-271}, \href{http://arxiv.org/abs/hep-th/0209214}{hep-th/0209214}.


\bibitem{AKSZ}M. Alexandrov, M. Kontsevich, A. Schwartz and O. Zaboronsky, ``\textit{The Geometry of the master equation and topological quantum field theory}", \href{http://dx.doi.org/10.1142/S0217751X97001031}{Int.J.Mod.Phys. A12 (1997) 1405-1430}, \href{http://arxiv.org/abs/hep-th/9502010}{arXiv:hep-th/9502010}.

\bibitem{BV1}
I. A. Batalin and G. A. Vilkovisky, ``\textit{Gauge Algebra and Quantization}", \href{http://dx.doi.org/10.1016/0370-2693(81)90205-7}{Phys.Lett. B102 (1981) 27-31}.

\bibitem{BV2}
I. A. Batalin and G. A. Vilkovisky, ``\textit{Quantization of Gauge Theories with Linearly Dependent Generators}", \href{http://dx.doi.org/10.1103/PhysRevD.30.508}{Phys.Rev. D28 (1983) 2567-2582}, \href{http://dx.doi.org/10.1103/PhysRevD.28.2567}{Phys.Rev. D30 (1984) 508}.

\bibitem{BV3}
I. A. Batalin and G. A. Vilkovisky, ``\textit{Existence Theorem for Gauge Algebra}", \href{http://dx.doi.org/10.1063/1.526780}{J.Math.Phys. 26 (1985) 172-184 }.

\bibitem{Boulanger2012}
N. Boulanger, N. Colombo and P. Sundell, ``\textit{A minimal BV action for Vasiliev's four-dimensional higher spin gravity}'', \href{http://dx.doi.org/10.1007/JHEP10(2012)043}{JHEP 1210 (2012) 043}, \href{http://arxiv.org/abs/arXiv:1205.3339}{arXiv:1205.3339 [hep-th]}.

\bibitem{Cesar}
C. Arias, P. Sundell and A. Torres--Gomez, ``\textit{Two-dimensional QP sigma models in vectorial superspaces}'', to appear.








\end{thebibliography}
\end{document}